\documentclass[fleqn,a4paper,10pt]{article}
\usepackage[utf8]{inputenc}
\usepackage[pdftex]{graphicx}
\usepackage{amsmath}
\usepackage{amssymb}
\usepackage[hmargin=2.5cm,vmargin=3.5cm]{geometry}
\usepackage{color}
\usepackage[usenames,dvipsnames]{xcolor}
\usepackage{wasysym}
\usepackage{enumerate}
\usepackage{setspace}
\usepackage{placeins}
\usepackage{multirow}
\usepackage{lineno}

\date{}

\title{The dynamics of copper intercalated molybdenum ditelluride}
\author{Nicolas Onofrio$^1$, David Guzman$^2$ and Alejandro Strachan$^2$\footnote{Corresponding author: strachan@purdue.edu}}

\begin{document}

\maketitle

\noindent
$^1$ Department of Applied Physics, The Hong Kong Polytechnic University, Hong Kong SAR \\
$^2$ School of Materials Engineering and Birck Nanotechnology Center Purdue University, West Lafayette, IN 47906 USA \\

\begin{abstract}
Layered transition metal dichalcogenides are emerging as key materials in nanoelectronics and energy applications.
Predictive models to understand their growth, thermomechanical properties and interactions with metals are needed in order to accelerate their incorporation into commercial products.
Interatomic potentials enable large-scale atomistic simulations at the device level, beyond the range of applications of first principle methods. 
We present a ReaxFF reactive force field to describe molybdenum ditelluride and its interactions with copper.
We optimized the force field parameters to describe the properties of layered MoTe$_2$ in various phases, the intercalation of Cu atoms and clusters within 
its van der Waals gap, including a proper description of energetics, charges and mechanical properties. The training set consists of an extensive set of 
first principle calculations computed from density functional theory.
We use the force field to study the adhesion of a single layer MoTe$_2$ on a Cu(111) surface and the results are in good agreement with density functional theory, even 
though such structures were not part of the training set.
We characterized the mobility of the Cu ions intercalated into MoTe$_2$ under the presence of an external electric fields via molecular dynamics simulations. The results show
a significant increase in drift velocity for electric fields of approximately 0.4 V/\AA~and that mobility increases with Cu ion concentration.

\end{abstract}

{\bf Keywords}: molecular dynamics, ReaxFF, MoTe$_2$, intercalation, TMD/metal interface

\section{Introduction}

The integration of single-layer and few-layers transition metal dichalcogenides (TMDs) into nanoelectronics~\cite{radisavljevic2011single} is paving the way for future developments 
of low power, ultrathin electronic devices~\cite{jariwala2014emerging,das2014toward}.
The versatile chemical composition and phase stability of TMDs results in a wide range of electronic properties from metallic to wide band gap semiconductor~\cite{chhowalla2013chemistry}.
Additionally, the two dimensional nature of TMDs enables its hybridization and integration with various materials via intercalation~\cite{morosan2006superconductivity,wang2015physical}, 
doping~\cite{komsa2012two,wang2015physical}, heterostructure~\cite{yu2015vertical,wang2015physical} and interfaces~\cite{allain2015electrical}.
Although the majority of efforts so far have been devoted to TMD-based transistors~\cite{fang2012high,liu2013role},  TMDs are rapidly finding other applications including metallic 
interconnects~\cite{lin2014flexible} (see also Cu-graphene hybrids~\cite{mehta2015enhanced}), non-volatile memory 
devices~\cite{bertolazzi2013nonvolatile,bessonov2015layered,tan2015non} and electro/mechanical switches have been predicted from theory~\cite{duerloo2014structural,li2016structural}.

An important limitation of TMD-based electronics lies in the poor hybridization between metals and TMDs resulting in high interface resistance~\cite{allain2015electrical}.
Therefore, in addition to the inherent Schottky barrier at the metal/semiconductor interface, a tunneling barrier arise from the van der Waals (vdW) nature of the metal-TMD interaction.
Moreover, the vdW interaction between multiple layers of TMDs also limits carrier injection at the interface with the metal contact.
Various metals have been investigated in order to improve contacts~\cite{gong2013metal} as well as local TMD processing such as intercalation, phase engineering~\cite{kappera2014phase} 
and chemical doping~\cite{fang2012high}.
Due of the wide compositional flexibility of TMDs and the various metals of interest, brute force experimental optimizations are likely to be ineffective and predictive
computational techniques can help narrow down experimental studies to the most promising candidates. 
Electronic structure calculations using density functional theory (DFT), including corrections for the nonlocal dispersive forces part of vdW interactions~\cite{grimme2010consistent,lee2010higher}, have
played and continue to play a key role in this field~\cite{liu2013role,kang2014high,kang2014computational,farmanbar2016first}. 
However, the computational intensity of DFT calculations has limited its application to relatively small size, model systems composed of few hundreds of atoms.
Thus, most of the DFT studies have focused on defect free TMD/metal interfaces (limited by the small size of the supercell) and little is known about the nanostructure 
of the interface. 

Direct synthesis of single to few layers TMDs have been demonstrated however, even with sophisticated chemical vapor deposition (CVD) techniques~\cite{van2013grains,yun2015synthesis}, 
TMDs present high defect density~\cite{jeong2015visualizing}.
Understanding defect formation in TMDs would not only enable the synthesis of large area, defect-free layers but would also provide the knowledge for defect-controlled TMD growth.
Structural defects in TMDs, including point defects and grain boundary influence its electronic, thermal and mechanical properties~\cite{voiry2013conducting,feng2015tuning,lin2016defect}.
Unfortunately, the atomistic mechanisms of nucleation and growth during the CVD process remain unclear.

Large-scale atomistic simulations would be critical to study the interactions between defects and interfaces and to provide atomistic insight into the process of synthesis and 
growth of these materials. This paper contributes to bridge the gap between first principle calculations and experiments via the development of the first reactive force field 
capable of describing a TMD, a metal and their interactions, specifically molybdenum ditelluride and Cu. We use the ReaxFF formalism that has shown the flexibility to describe 
a wide range of materials including metals, semiconductors, their oxides~\cite{van2003reaxffsio,nielson2005development} and molecular materials~\cite{wood2014coupled,wood2015nonequilibrium,wood2015ultrafast}. 
The force field is parameterized to describe various phases of Cu, Mo and MoTe$_2$ as well as the corresponding interfaces, the intercalation of Cu atoms and small 
clusters in the vdW gap of multiple layers of MoTe$_2$ including a proper description of the relative energies, charges and mechanical properties.

The rest of the paper is organized as follows. In Section \ref{sec:method} we present details about the simulations and methods. Section \ref{sec:reaxff} describes the optimized 
force field and presents its comparison with DFT calculations. In Sections \ref{sec:app1} and \ref{sec:app2} we study the interface between monolayer MoTe$_2$ and a Copper 
slab and the dynamics of Cu-intercalated MoTe$_2$ under various external voltages, respectively. We finally draw conclusions in Section \ref{sec:conc}. 

\section{Force field optimization approach and methods}
\label{sec:method}

ReaxFF is a bond order interatomic potential and naturally allows bond breaking and forming during molecular dynamics (MD) simulations~\cite{van2001reaxff}. 
The total energy is defined as the sum of various contributions including covalent (bond stretch, angles, torsions, and over/under coordination) and non-bonded 
interactions (van der Waals and Coulomb); additional terms to account for lone pair and bond conjugation.
Electrostatic interactions are described using environment-dependent partial atomic charges which are computed at each step of the simulation using the charge equilibration 
method~\cite{mortier1986electronegativity,rappe1991charge} (QEq). 
All contributions to the covalent energy depend on the partial bond order between atoms, a function that smoothly decays to zero when the bond between atoms is broken. 
Details on the mathematical form of ReaxFF can be found in the original paper, Ref.~\cite{van2001reaxff} and a recent review of applications in Ref.~\cite{senftle2016reaxff}.
In order to determine the parameters that describe the functional form of the interaction, the force field has to be optimized against a training set consisting of first principle calculations
or a combination between first principles calculations and experiments. The parameterization of such a force field is complex due to the large number of parameters and the need
for extensive training sets and time-consuming optimization cycles. Thus, despite the enormous progress to date~\cite{senftle2016reaxff}, several materials of technological
importance lack accurate force field descriptions.

\subsection{DFT simulation details}

All DFT calculations have been performed with the Vienna Ab initio Simulation Package (VASP)~\cite{kresse1996efficient,kresse1996efficiency} within the generalized gradient 
approximation as proposed by Perdew, Burke, and Ernzerhof (PBE)~\cite{perdew1996generalized}. In addition, we use the non-local vdW density functional of Langreth and 
Lundqvist~\cite{dion2004van,klimevs2011van} (vdW-DF) throughout the paper to correct for the London dispersion, poorly described by PBE functionals.
Atomic structures have been relaxed until energies and forces reach a tolerance of 1$\times$10$^{-4}$ eV and 1$\times$10$^{-2}$ eV/\AA, respectively.
The kinetic energy cutoff for the plane-wave basis was set to 500 eV and we integrate the k-mesh with various number of points depending on the size of the supercell.
We use a dense 8$\times$8$\times$8 k-mesh for bulk Cu and Mo, an intermediate 4$\times$4$\times$2 grid for bulk MoTe$_2$ and Cu-intercalated MoTe$_2$ and gamma 
point calculations have been performed on large cells used to compute Cu mobility.
We verified that the kinetic energy cutoff chosen provides well converged structures with respect to energy, force and stress.
Diffusion of Cu ions in the vdW gap of MoTe$_2$ and minimum energy path between H $\rightarrow$ T' phases of MoTe$_2$ have been computed using the nudged elastic 
band~\cite{henkelman2000climbing} (NEB) method, as implemented in VASP.
Additionally, we performed {\it ab initio} MD simulations at 500 K with a kinetic energy cutoff of 300 eV and timestep of 1.5 fs in order to overcome the energy barrier corresponding 
to the dissociation of a Cu$_3$ cluster intercalated in MoTe$_2$.

\subsection{Parameter optimization}

In order to optimize the force field we use an in-house code written in Python that implements a Monte Carlo-based (MC) simulated annealing algorithm~\cite{kirkpatrick1984optimization}
and coupled to the LAMMPS simulator~\cite{plimpton1995fast} (via its Python interface) to evaluate energies, forces, pressures and charges of molecules and crystals included in 
the training set. This method has been used previously to optimize force fields for oxides~\cite{strachan1999phase} and metals~\cite{strachan2004first}.
The training set is included as an XML database of DFT calculations describing the dissociation of small molecules and equation of states for crystals.
The goal of the simulated annealing optimization is to minimize a total error function ($\epsilon_{tot}$) defined as the sum of individual discrepancies between the force field and DFT 
data, corresponding to energy (E), force (f), pressure (P) and charge (Q) as:

\begin{equation}
\epsilon_{tot} = \sum_{X=E,f,P,Q}\frac{\left(X^{ReaxFF}-X^{DFT}\right)^2}{N_X^2}
\end{equation}

with $N_X$ a set of normalization constants chosen such that each individual error appears with equivalent weight; these parameters are also
used to establish the relative importance between different quantities in the training set.

Before starting the simulated annealing run we perform an initial sensitivity analysis for each parameter in the force field in order to determine the
steps by which each parameter will be stochastically modified during the MC run. A Metropolis criterion allows the random selection of some ``non-optimum'' 
parameters in order to expand the exploration space and avoid the force field function becoming trapped in local minima. The {\it temperature} used
for the Metropolis acceptance is decreased during the simulation in order to converge towards the global minimum.
We note that the QEq parameters are optimized first in order to reproduce partial atomic charges of the structures in the training set obtained from Bader analysis~\cite{henkelman2006fast}
of the DFT runs. This is done by setting $\epsilon_E$, $\epsilon_f$ and $\epsilon_P$ to zero in the energy expression so that only partial charges are compared.
The simulated annealing code distributes MD simulations in parralel (using IPython for parallel computing) and at its current stage of development can perform 1,000 single point calculations in 
approximatively 5 seconds over 64 cores.

\section{Cu-MoTe$_2$ reactive force field}
\label{sec:reaxff}

MoTe$_2$ has recently received significant attention because of the discovery of a metallic T' phase~\cite{qian2014quantum,keum2015bandgap} 
energetically close~\cite{duerloo2014structural,li2016structural} to its ground state semiconducting H phase, common to group VI TMDs.
Therefore, we decided to develop parameters for MoTe$_2$ rather than the more studied molybdenum disulfide (MoS$_2$), already explored by 
MD~\cite{liang2009parametrization}. Copper is the metal of choice for electrodes and metallic interconnects; hence, we studied the system Cu-MoTe$_2$.
The following subsections describe the training data used in the force field parameterization and compare the optimized ReaxFF with the
DFT training data. 

\subsection{Initial parameters and charge equilibration}

We built an initial force field with parameters for Mo and Te extracted from Ref.~\cite{chenoweth2009reaxff} and Cu from Ref.~\cite{van2010development}.
The general parameters were re-optimized in order to improve the atomic features including equation of state and formation energy, as described
in details in the following sections. Therefore, the parameters presented here are not transferable to the initial ReaxFF they belong to and the new force field
we propose corresponds to a new branch of the ReaxFF tree, as defined in Ref.~\cite{senftle2016reaxff}. Additionally, we introduce three anglar terms 
corresponding to Te-Mo-Te, Te-Cu-Te and Cu-Te-Cu in order to describe the subtle energy difference between MoTe$_2$ phases and the potential energy 
landscape of Cu diffusion in MoTe$_2$.
We first optimized the QEq parameters, traditionally named $\gamma_{EEM}, \chi_{EEM}$ and $\eta_{EEM}$ representing the shielding distance for Coulomb
interactions, electronegativity and hardness parameters, respectively.
Partial atomic charges were optimized on each atom against Bader charges.
We show in Table \ref{tab:tab1} the root mean square error of charges per atom computed between the optimized ReaxFF and DFT.
ReaxFF describes accurately charges with an average error of 0.037 $\pm$ 0.015 $e$ per atom compared to Bader charges.

\subsection{Bulk phases and bond dissociation}

\begin{figure}[!ht]
  \centering
  (a)\includegraphics[width=0.65\textwidth]{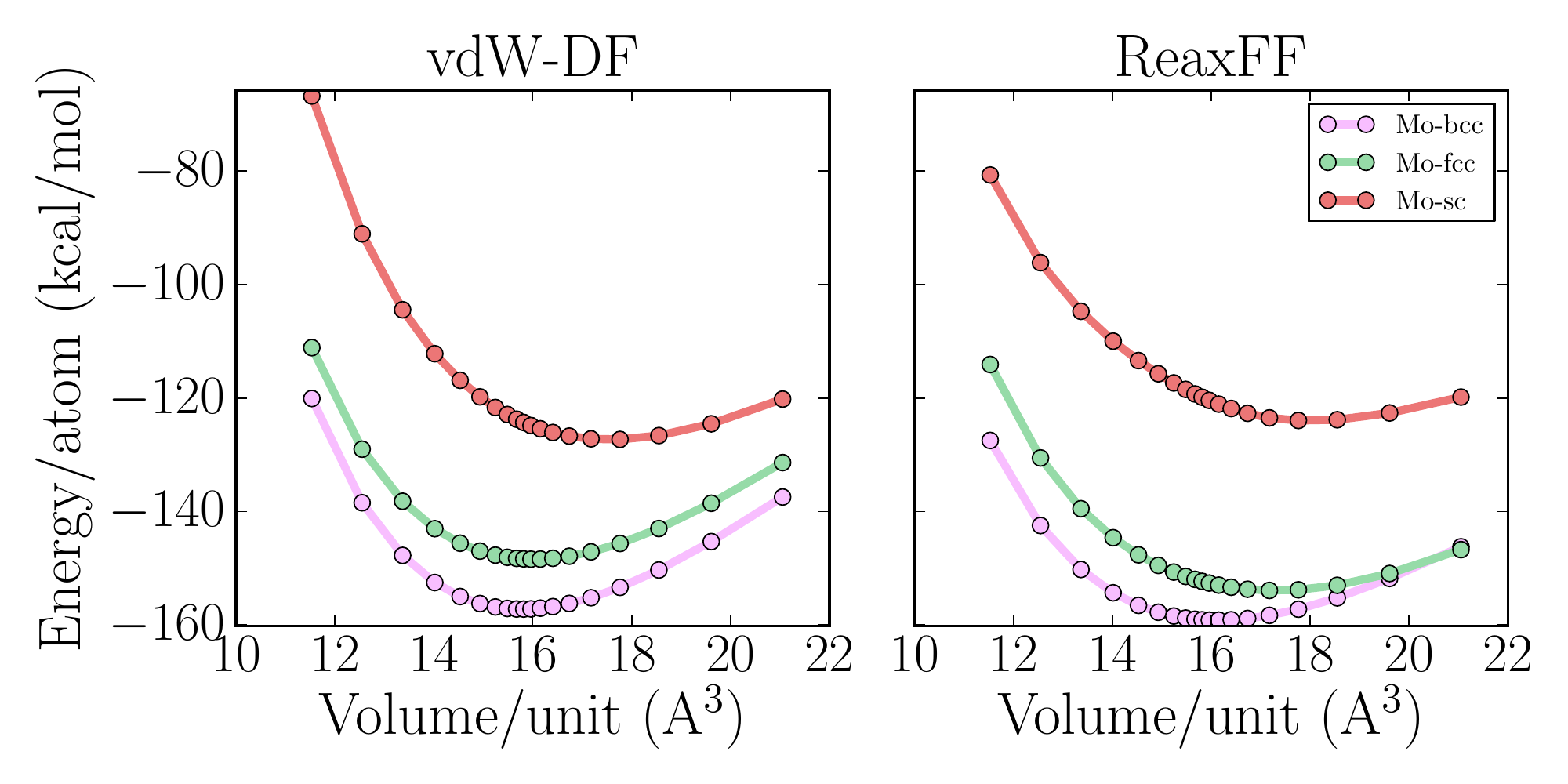}
  (b)\includegraphics[width=0.65\textwidth]{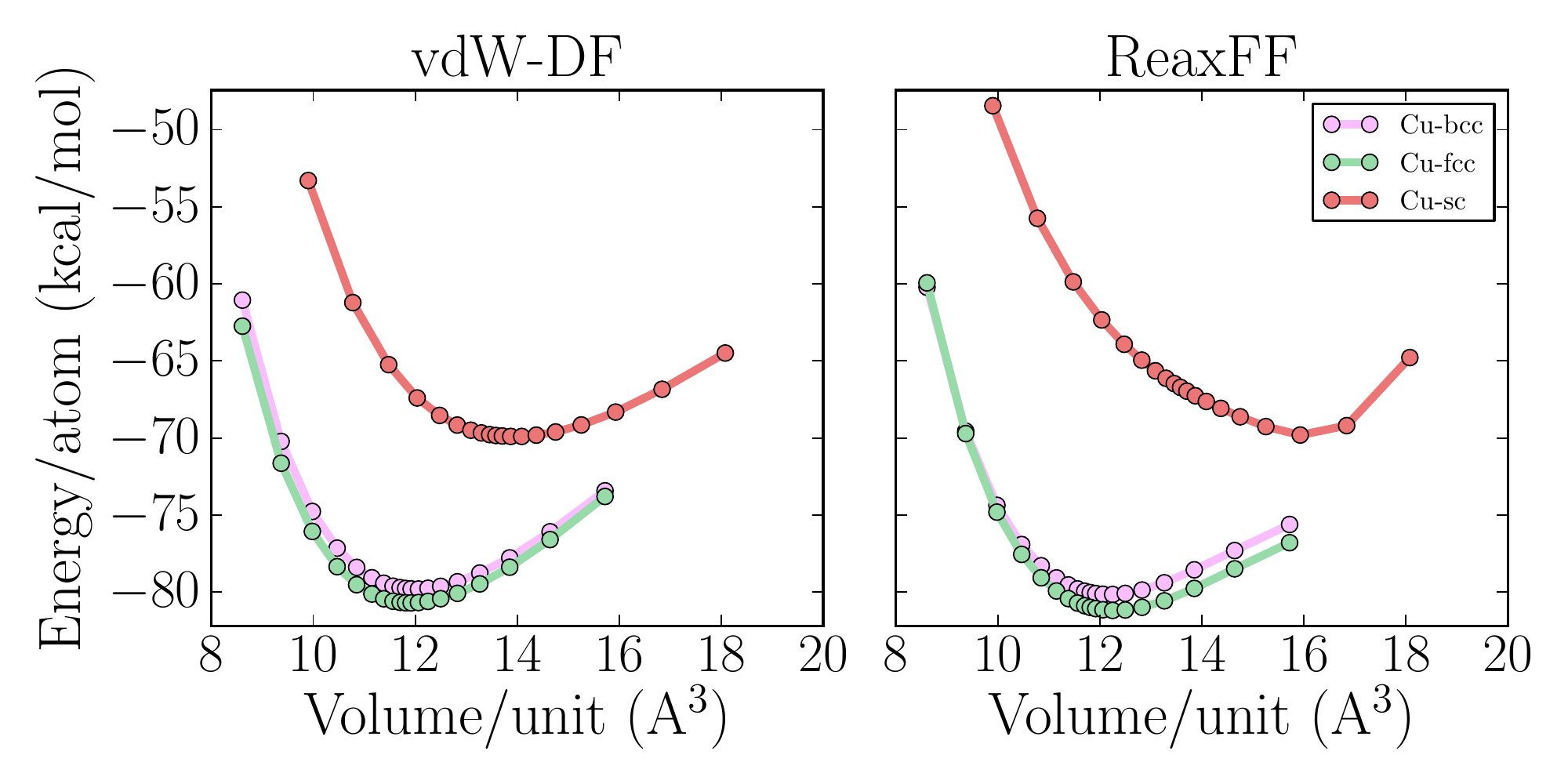}
  \caption{Equation of states for various bulk phases of Mo (a) and Cu (b) computed with vdW-DF (left) and ReaxFF (right) as a function of the volume of the unit cell.
  The minimum energy of the EOS have been shifted to the enthalpy of formation of the corresponding crystal.}
  \label{fig:fig1}
\end{figure}

A key feature of ReaxFF lies in its ability to describe atoms in various chemical environments, corresponding to different coordinations.
Therefore, we fit ReaxFF against various bulk phases of Cu and Mo including simple cubic (SC), face centered cubic (FCC) and body centered cubic (BCC). 
Figure \ref{fig:fig1} shows various equations of state (EOS) computed with DFT and ReaxFF, with minimum energy shifted to their corresponding enthalpies 
of formation (see Table \ref{tab:tab1} for details). For both Mo and Cu, ReaxFF describes with accuracy the ground state structures, only the simple cubic phases 
present some discrepancy with respect to DFT.
However, the simple cubic phases lie tens of kcal/mol higher than the ground state and therefore do not represent important contribution to their chemistry;
it is only important that they remain energetically prohibited.

\begin{figure}[!ht]
  \centering
  \includegraphics[width=0.65\textwidth]{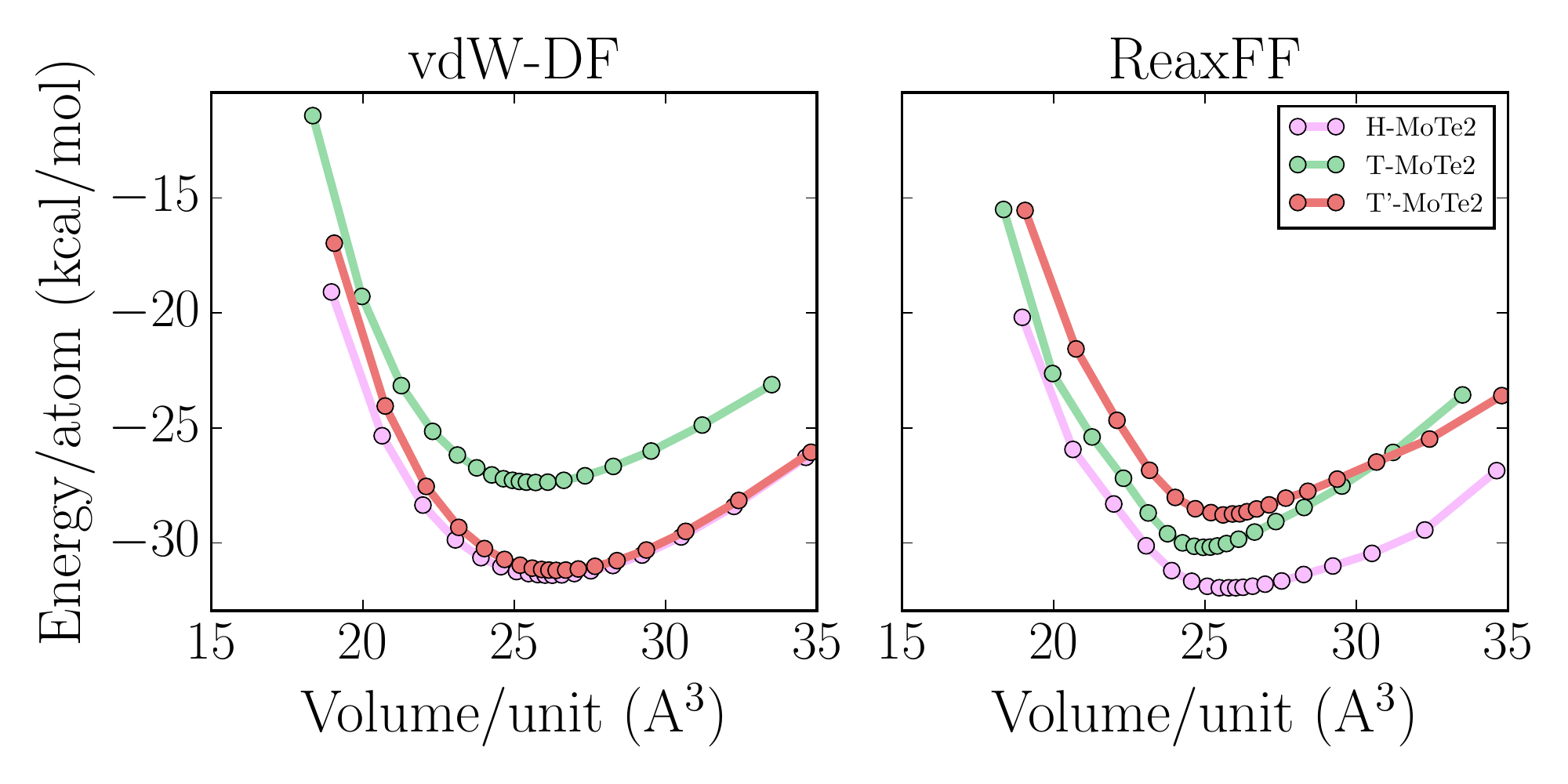}
  \caption{Equation of states for various bulk phases of MoTe$_2$ computed with vdW-DF (left) and ReaxFF (right) as a function of the volume of the unit cell.
  The minimum energy of EOS have been shifted to the binding energy of the corresponding phase.}
  \label{fig:fig2}
\end{figure}

\begin{figure}[!ht]
  \centering
  \includegraphics[width=0.45\textwidth]{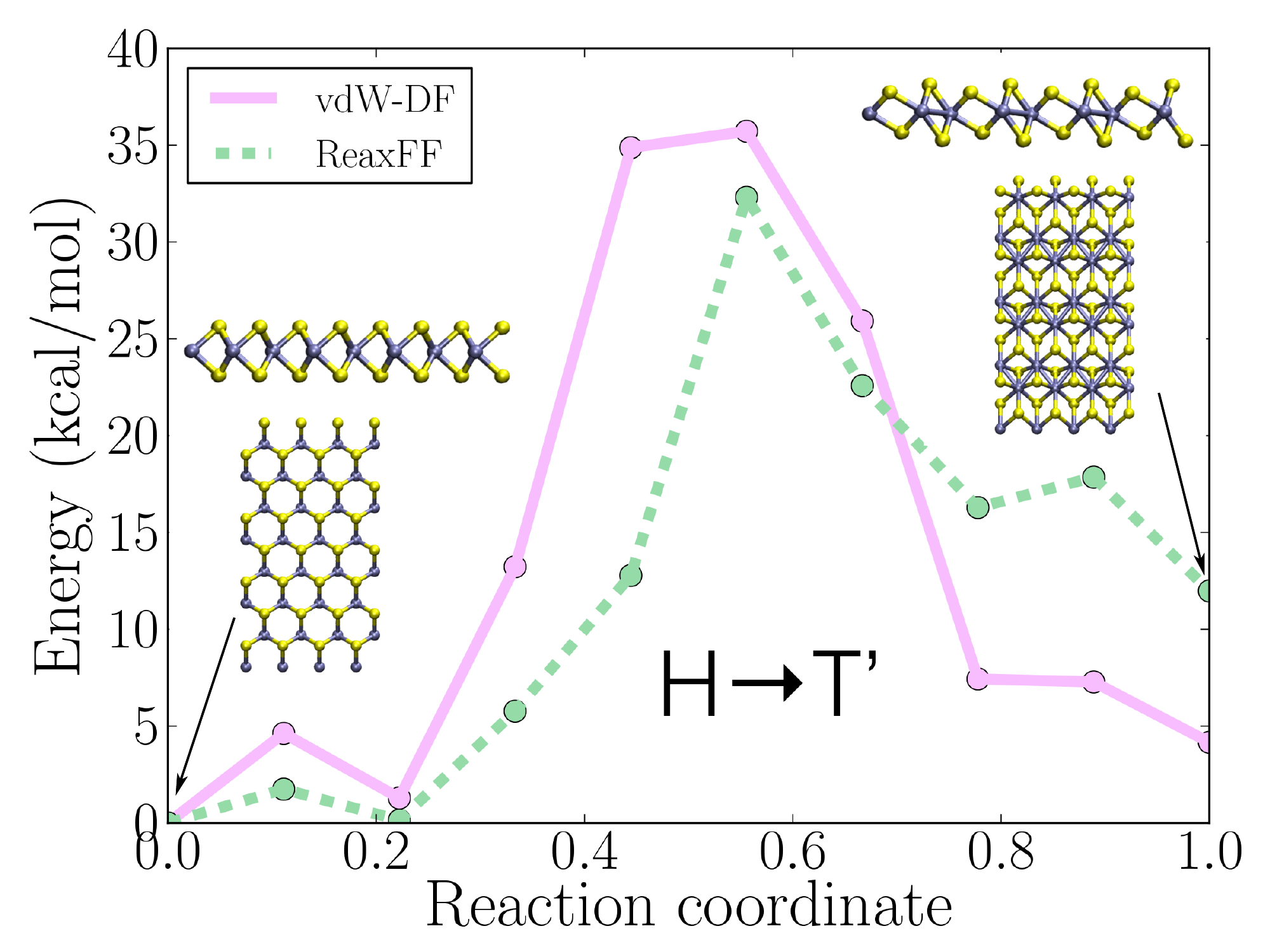}
  \caption{Potential energy surface corresponding to the transition of rectangular monolayer MoTe$_2$ from the phase H $\rightarrow$ T' computed with vdW-DF (solid) 
  and ReaxFF (dashed) as a function of reaction coordinate.}
  \label{fig:fig3}
\end{figure}

MoTe$_2$ has been observed in the stable H and T' phases as well as a metastable T phase, corresponding to trigonal prismatic, distorted octahedral and octahedral 
geometries, respectively. The semiconducting H phase has been shown to be the ground state configuration for Mo-based TMDs while the metallic T' phase lies few 
meV above the H for MoTe$_2$. A transition from H to T' has been predicted from DFT calculations by mechanical strain in the range of 0.3 to 3\%~\cite{duerloo2014structural} or an applied electrostatic gate voltage~\cite{li2016structural}.
Figure \ref{fig:fig2} shows the EOS corresponding to various phases of bulk MoTe$_2$ computed with DFT and ReaxFF, and shifted to their binding energy.
The binding energy is defined as the total energy of the MoTe$_2$ phase minus the total energies of bulk Mo in its ground state BCC phase and the diatomic molecule Te$_2$.
The force field predicts the H phase to be the ground state however, the T' phase appears higher than the T phase.
Additionally, we computed the potential energy surface corresponding to transition between H$\rightarrow$T' within the rectangular unit cell constrained to the initial H phase, as
depicted in Figure \ref{fig:fig3}. We found that ReaxFF predicts the energy barrier between H and T' phases E$_A$=32.3 kcal/mol in good agreement compared to 35.7 kcal/mol computed
from DFT calculations. We note that ReaxFF is able to capture subtle energy differences between the various phases of MoTe$_2$.

\begin{figure}[!ht]
  \centering
  \includegraphics[width=0.45\textwidth]{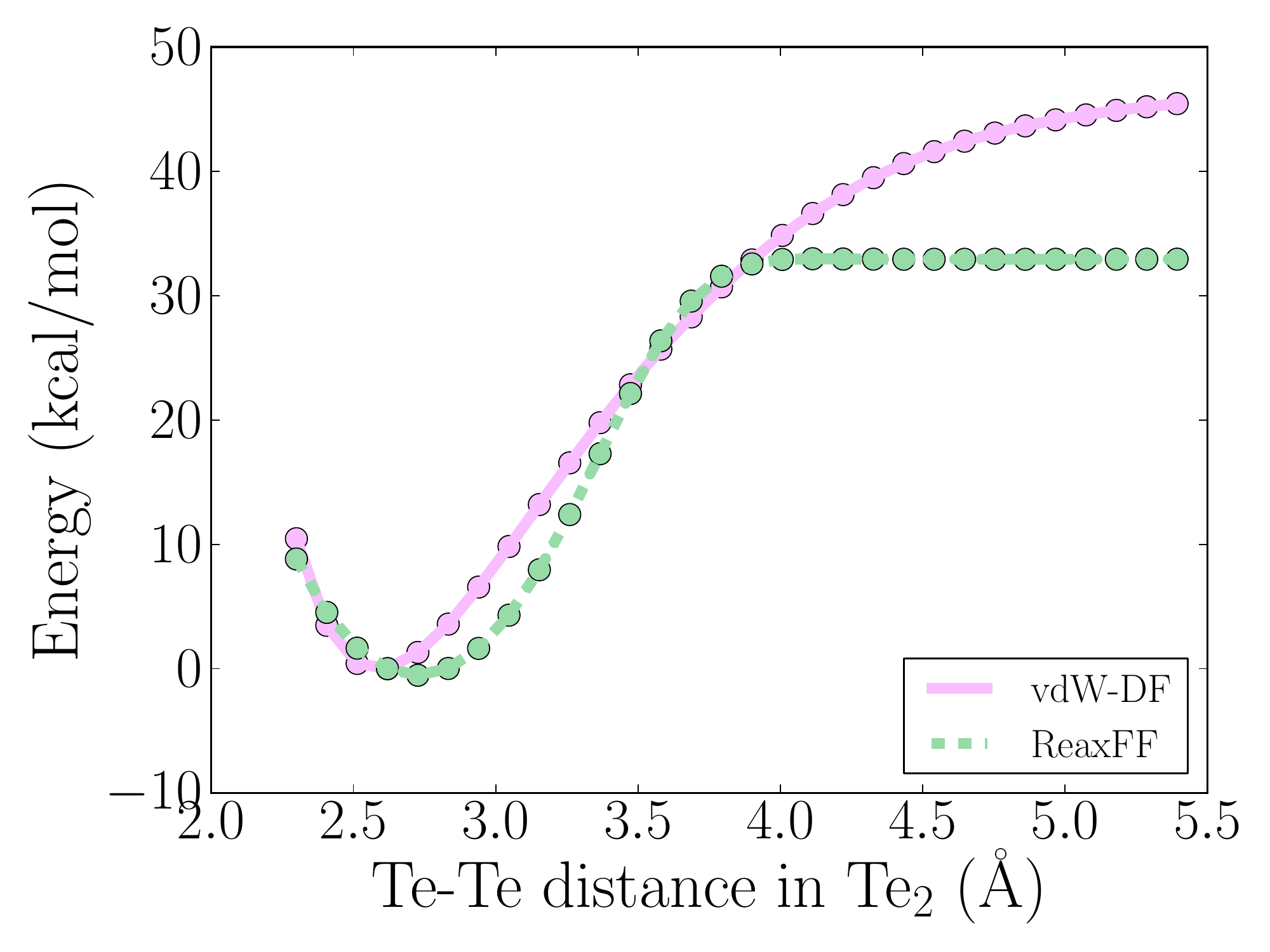}
  \includegraphics[width=0.45\textwidth]{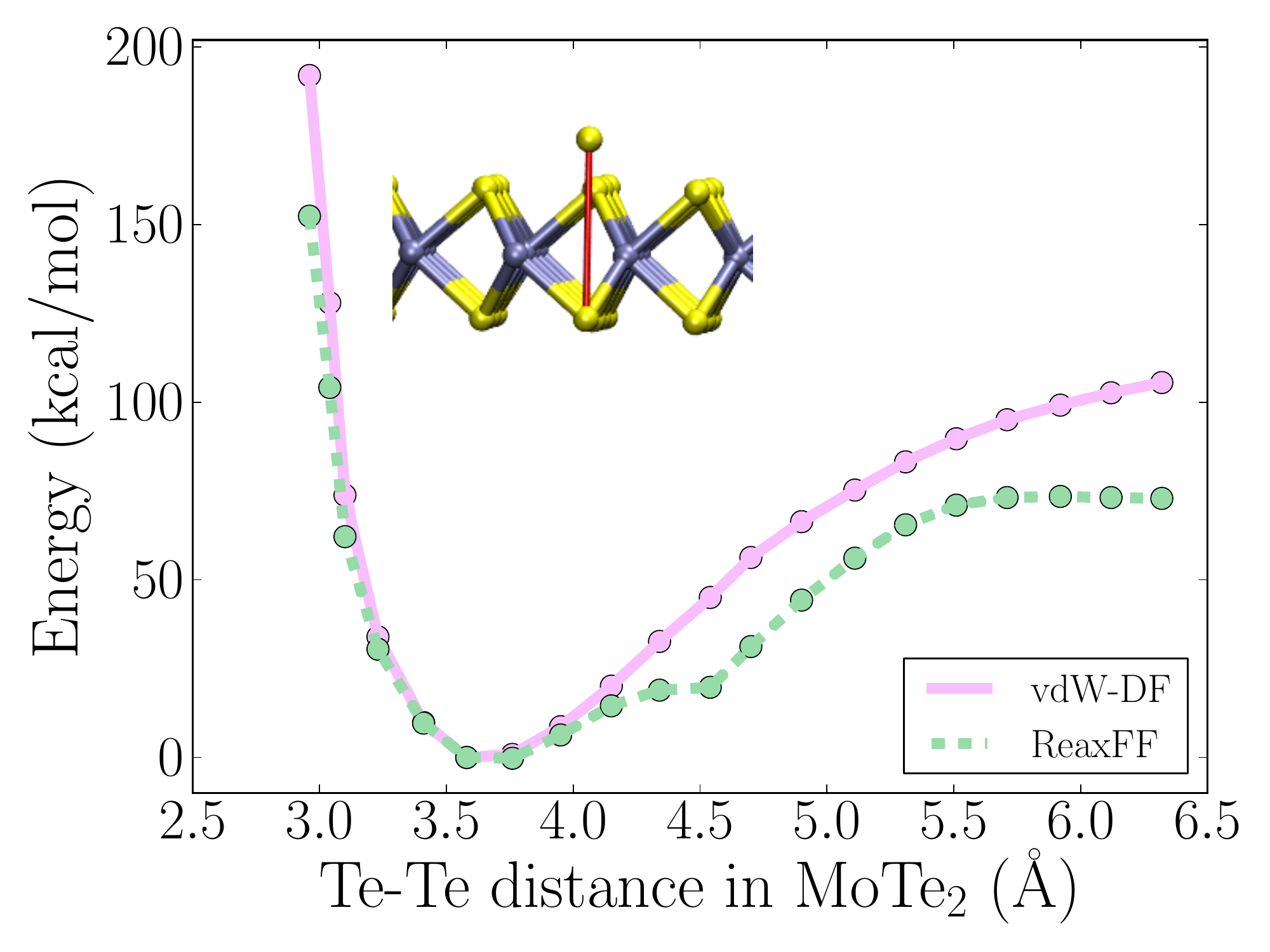}
  \caption{Dissociation curve of the Te$_2$ molecule (left) and Te atom from $4\times 4$ monolayer MoTe$_2$ (left) computed with vdW-DF (solid) and ReaxFF (dashed) as a function 
  of the corresponding bond distance.}
  \label{fig:fig4}
\end{figure}

The bond order nature of ReaxFF enables dynamical connectivity during molecular dynamics simulations and the ability to describe dissociation (and formation) of chemical bonds.
Therefore, the training set includes various bond dissociation curves, such as Te-Te and Mo-Te bonds in Te$_2$ molecule and monolayer MoTe$_2$, respectively.
The corresponding energies versus bond distance plots are shown in Figure \ref{fig:fig4}. We found a good agreement between ReaxFF and DFT even though the force field 
predicts slightly lower dissociation energies for both Te-Te and Mo-Te dissociations.

Finally, in order to validate the stability of the force field we performed molecular dynamics simulations of bulk and monolayer MoTe$_2$ in the microcanonical ensemble with a timestep of 0.5 fs.
We found little drift in the total energy ($\sim$ 1$\times$10$^{-6}$ kcal/mol/ps/atom) suggesting a good stability of the force field.

\subsection{Cu-intercalated MoTe$_2$}

\begin{figure}[!ht]
  \centering
  (a)\includegraphics[width=0.65\textwidth]{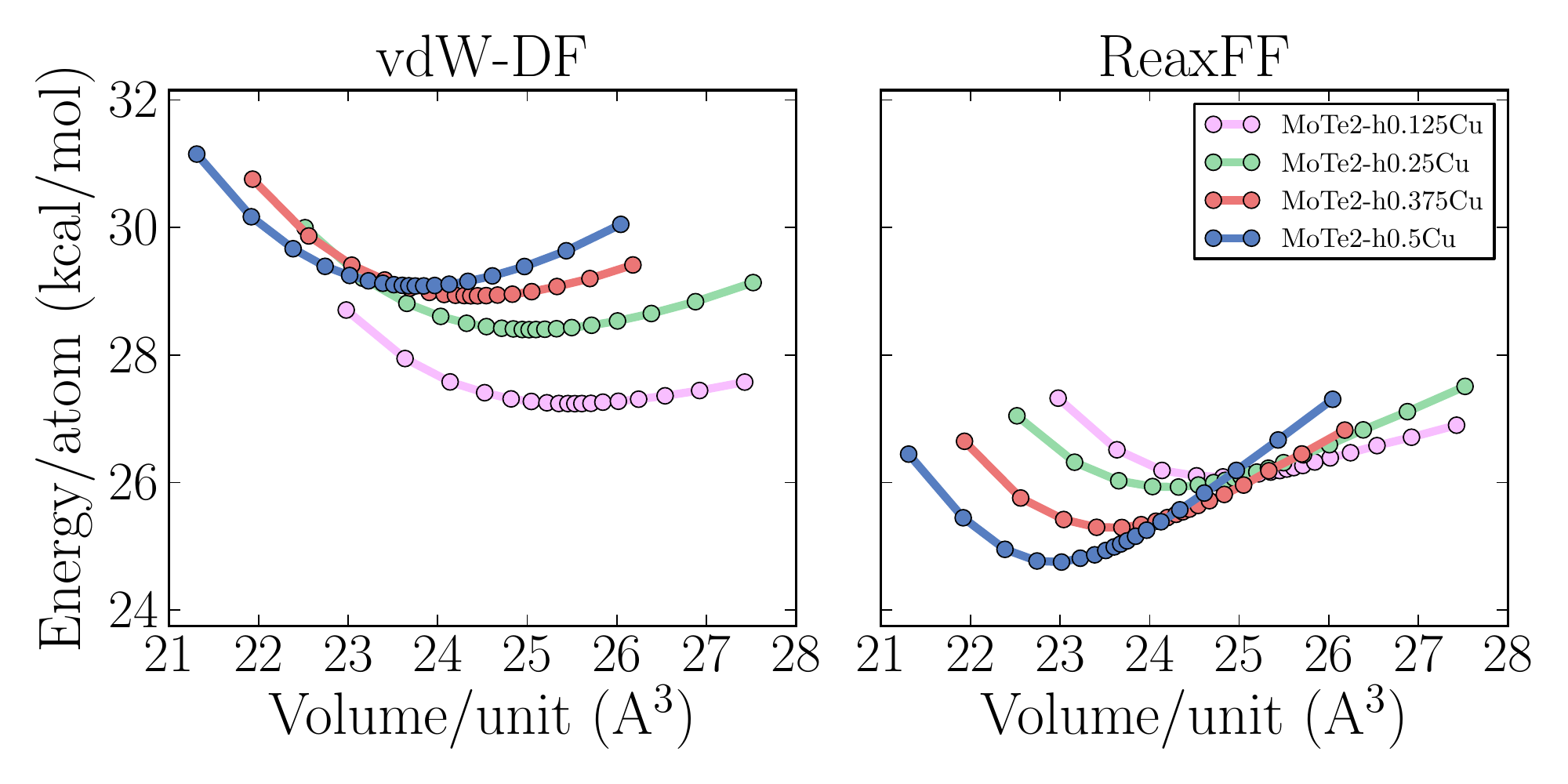}
  (b)\includegraphics[width=0.65\textwidth]{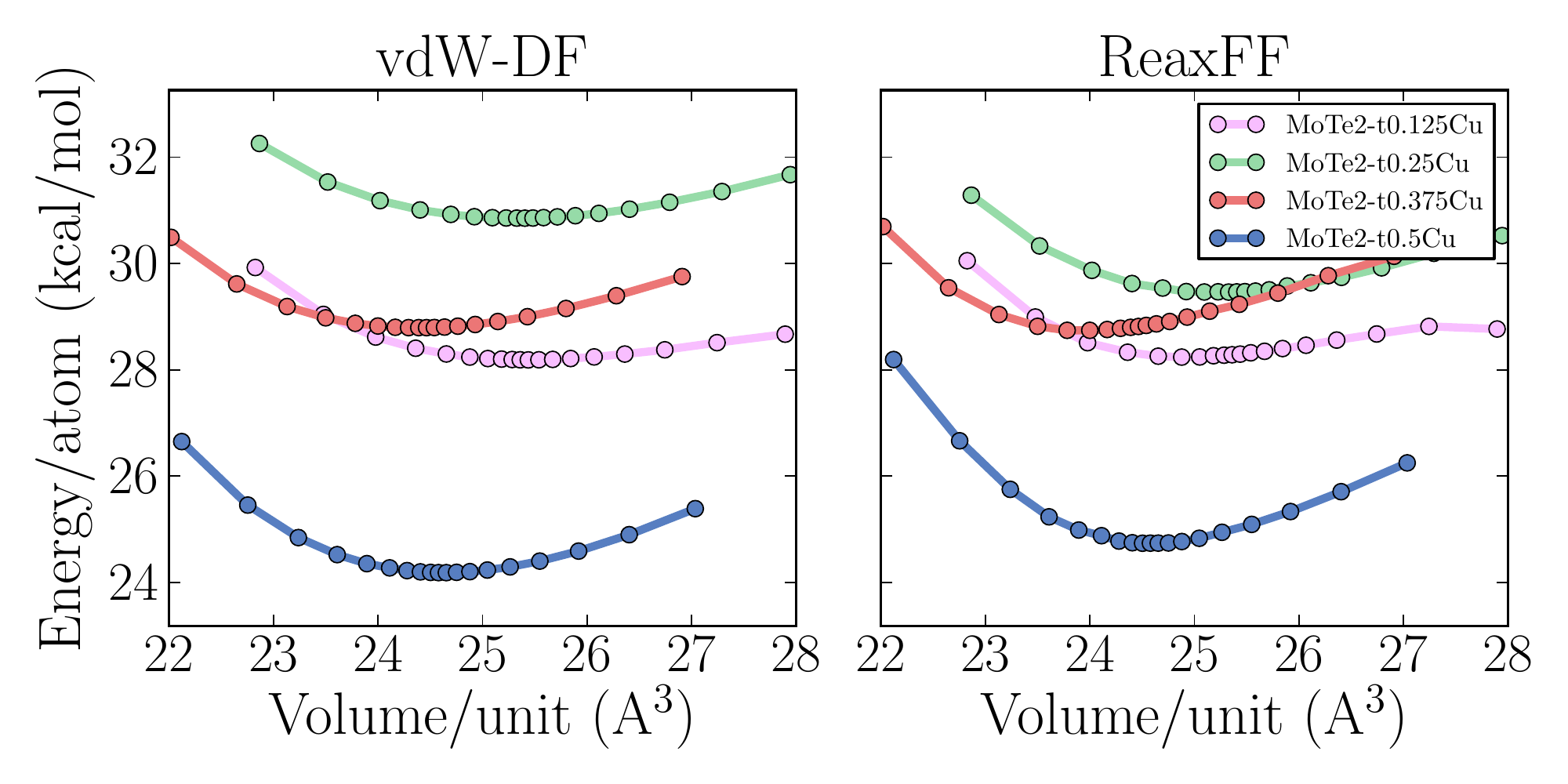}
  \caption{Energies corresponding to uniaxial deformation along the axis perpendicular to the MoTe$_2$ layer for various concentration of Cu intercalated at $h$- and $t$-sites 
  computed with vdW-DF (left) and ReaxFF (right) as a function of the volume of the unit cell.}
  \label{fig:fig5}
\end{figure}

The intercalation of Cu atoms within the vdW gaps of ground state H-MoTe$_2$ can occur either at octahedral (or hollow, noted ``h'' ) or tetrahedral (``t'') sites.
Figure \ref{fig:fig5} shows the energetics of $h$- (Fig. \ref{fig:fig5}a) and $t$- (Fig. \ref{fig:fig5}b) intercalated Cu$_x$MoTe$_2$ for Cu concentrations 
$x$=0.125, 0.25, 0.375 and 0.5. We show energy as a function of volume upon uniaxial deformation along the direction perpendicular to the plane of the TMD in order
to better test the interactions between Cu and the TMDs.
The energy curves in Figure \ref{fig:fig5} are referenced to the perfect TMD and Cu as: $E_f^{Cu} = E_{Cu_xMoTe_2}-E_{MoTe_2}-xE_{Cu}$ where $E_{Cu_xMoTe_2}$, 
$E_{MoTe_2}$ and $E_{Cu}$ are the energies of Cu$_x$MoTe$_2$, bulk MoTe$_2$ and Cu in its ground state FCC phase, respectively. 

The overall ReaxFF description of the structures and energetics is very accurate. The force field predicts the formation energy of the low concentration Cu h-intercalated 
MoTe$_2$ E$_f^{0.125h}$=26.2 kcal/mol, lower than the corresponding t-intercalated energy E$_f^{0.125t}$=28.3 kcal/mol, consistent with the predicted DFT energies 
of 27.2 and 28.2 kcal/mol, respectively.
However, our force field does not describe with accuracy the trends of formation energy corresponding to $h$-intercalated Cu-MoTe$_2$ systems.
The intercalation of Cu in h-sites MoTe$_2$ requires a narrow range of energies between 27.2 to 29.1 kcal/mol, which is challenging to be resolved by the force field.

\begin{figure}[!ht]
  \centering
  \includegraphics[width=0.45\textwidth]{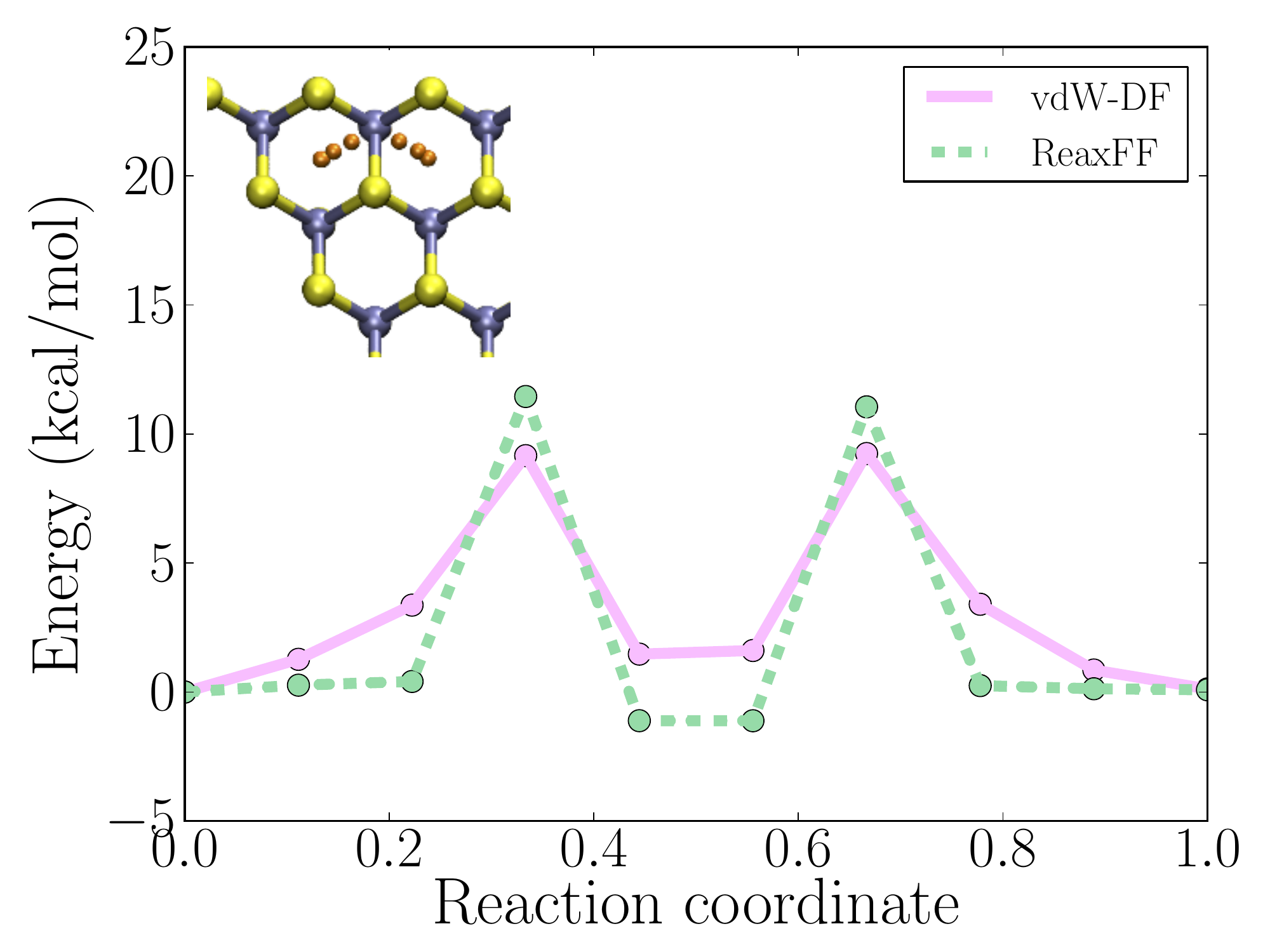}
  \includegraphics[width=0.45\textwidth]{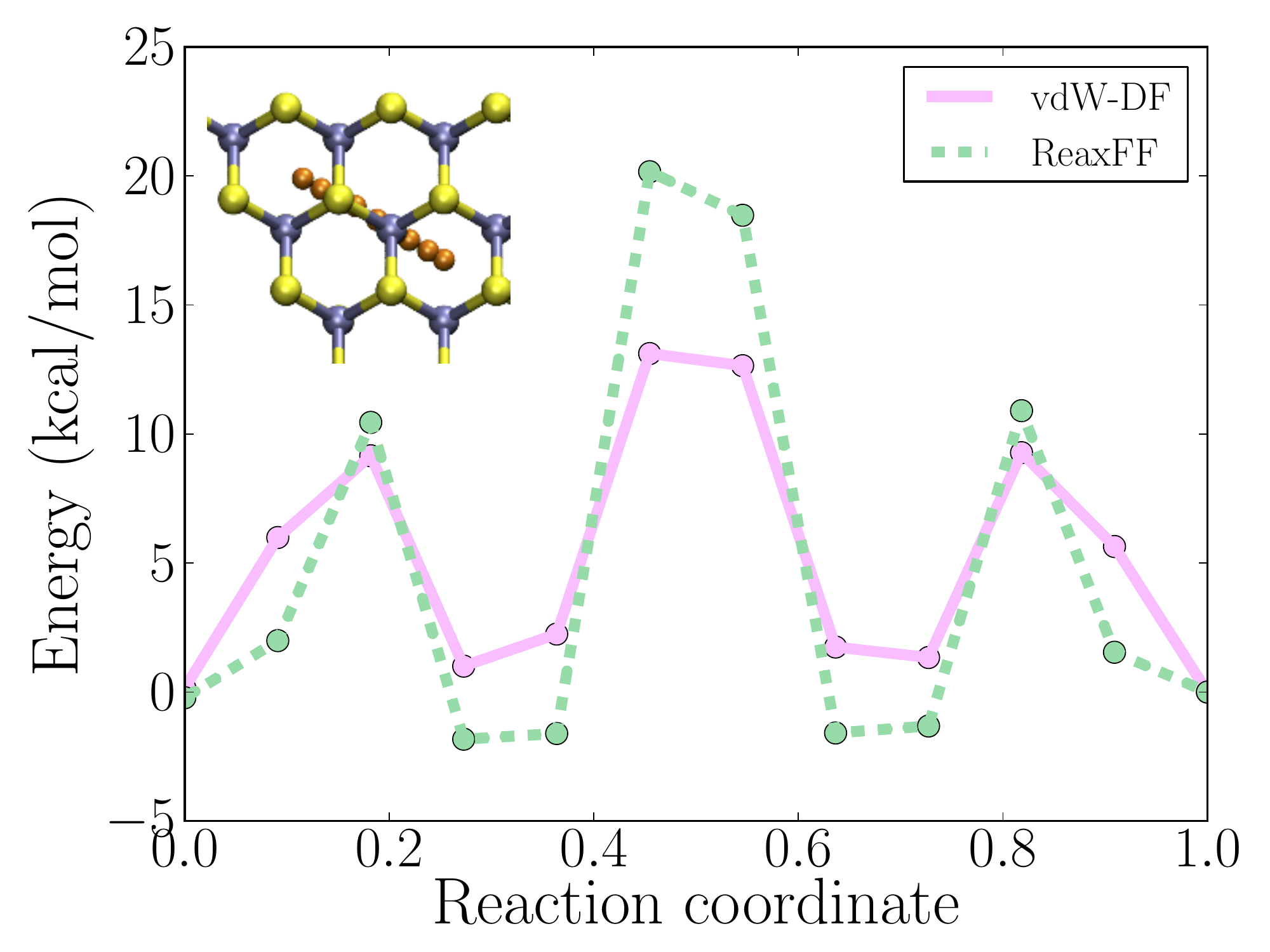}
  \caption{Potential energy surface corresponding to the migration of a copper atom within the vdW gap of $4\times 4$ bulk MoTe$_2$ computed with vdW-DF (solid) 
  and ReaxFF (dashed) as a function of reaction coordinate. Cu diffusion between neighboring octahedral sites through a tetrahedral site (left) and Cu diffusion
  between 2$^{nd}$ neighbor octahedral sites via two tetrahedral sites.}
  \label{fig:fig6}
\end{figure}

Figure \ref{fig:fig6} shows the potential energy surface corresponding to migration of Cu between neighboring octahedral sites as well as 2$^{nd}$ neighbor octahedral sites in 
bulk MoTe$_2$. From DFT calculations, we found the energy barriers for Cu diffusion from octahedral to tetrahedral and between two tetrahedral sites E$_A^{h\rightarrow t}$=9.2 
kcal/mol and E$_A^{t\rightarrow t}$=13.1 kcal/mol, respectively. The force field slightly over estimates the energy barriers E$_A^{h\rightarrow t}$=11.5 kcal/mol and 
E$_A^{t\rightarrow t}$=20.2 kcal/mol.

\begin{figure}[!ht]
  \centering
  \includegraphics[width=0.45\textwidth]{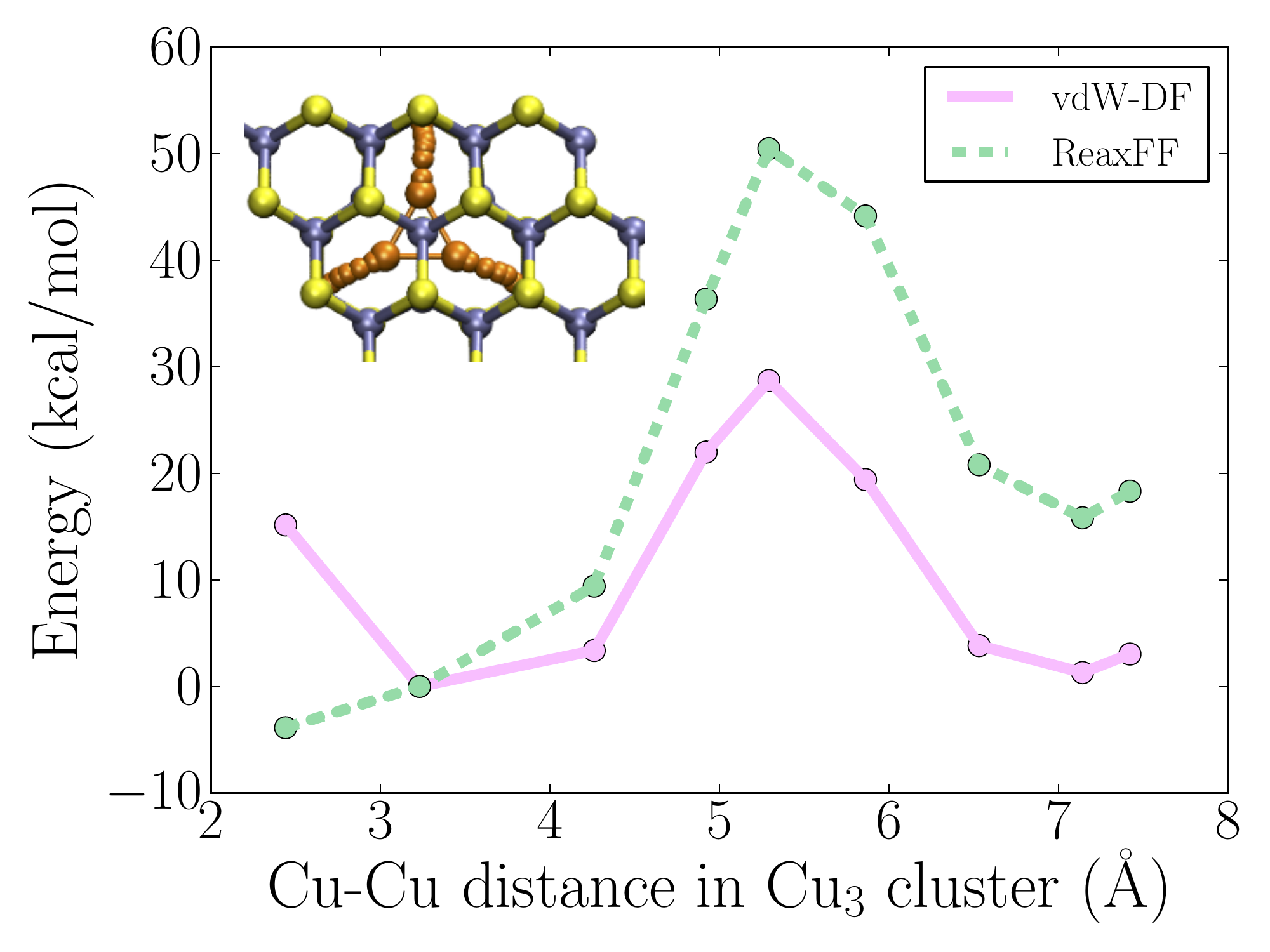}
  \includegraphics[width=0.45\textwidth]{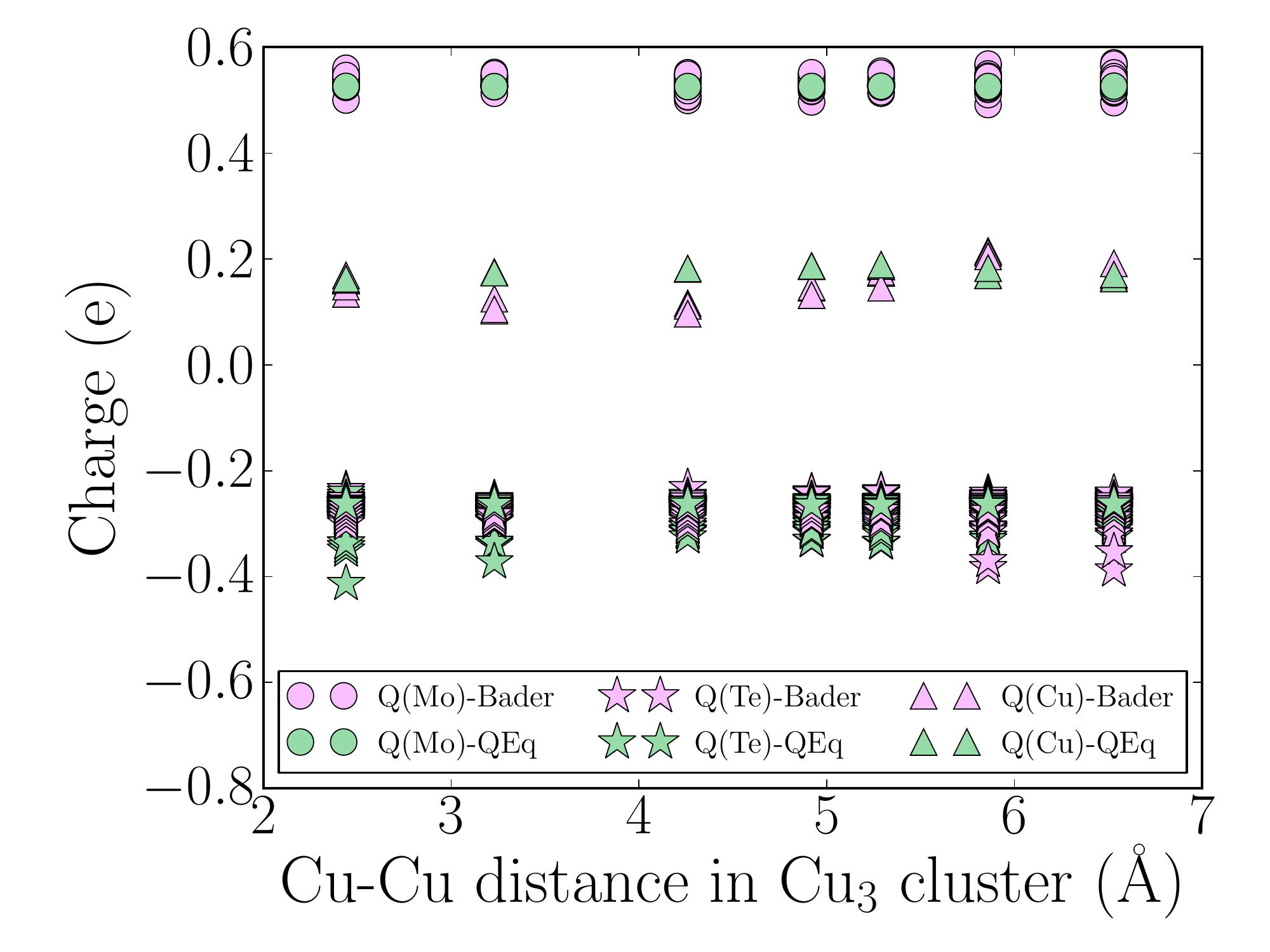}
  \caption{Potential energy surface corresponding to the dissociation of a copper cluster Cu$_3$ intercalated inside the vdW gap of $4\times 4$ bulk MoTe$_2$ computed with vdW-DF and ReaxFF (left) 
  and the corresponding Bader and QEq partial charges (right) as a function of the average Cu-Cu distance in the Cu$_3$ cluster.}
  \label{fig:fig7}
\end{figure}

In order to capture the proper dynamics of Cu diffusion and clusterization inside the vdW gap of MoTe$_2$, we performed {\it ab initio} MD simulations of Cu$_3$ cluster intercalated MoTe$_2$.
As shown in the inset of Figure \ref{fig:fig7}, we start the MD simulation with a Cu$_3$ cluster located around a tetragonal site with each atom in the cluster seating in the corner of an octahedral site.
MD simulation at 500 K shows the dissociation of the Cu$_3$ cluster into three isolated Cu atoms moving away from each other toward an opposite tetrahedral site.
Nine snapshots along the dissociation have been relaxed with DFT constraining the $xy$ (in-plane) dimensions of the Cu atoms and we extracted the potential energy surface presented in Figure \ref{fig:fig7}.
DFT calculations predict that Cu$_3$ needs to overcome an energy barrier $E_A$=28.1 kcal/mol for dissociation. 
On the other hand, ReaxFF overestimates the barrier by a factor of 1.75 leading to an activation energy $E_A$=49.4 kcal/mol.
Moreover, we compare on Figure \ref{fig:fig7} (right panel) the charges on each atom along the dissociation of the Cu$_3$ cluster computed with Bader analysis and ReaxFF's QEq.
Charges on each atom are roughly constant and we found a very good agreement between charges computed with the force field and DFT.

To summarize, we report on Table \ref{tab:tab1} the root mean square error per atom corresponding to charges, forces and pressure computed between DFT and ReaxFF as well as the 
various formation energies. We acknowledge that the force field presents some uncertainties, and the main deviations, with respect to DFT calculations, are related to the energy barriers 
for Cu diffusion and cluster dissociation. The overestimation of diffusion barriers suggest a slower mobility of Cu atoms described by ReaxFF than predicted from DFT.
We provide the optimized ReaxFF parameters in Supplementary Materials.

\begin{table}[!ht]
\caption{Root mean square error per atom (Err.) between the DFT (vdW-DF) and ReaxFF charges (Q), forces (f) and pressures (P) as well as formation energies (E$_f$), 
equilibrium volumes (V$_0$) and elastic properties (bulk moduli B$_0$ and elastic constants C$_{11}$) for various compounds included in the training set.}
\begin{center}
\begin{tabular}{c|ccc|ccc}
\multirow{3}{*}{Crystal/molecule} & Err. Q$^\ast$ & Err. f$^\ast$ & Err. P$^\ast$ &  E$_f^\star$ & V$_0^\dagger$ & B$_0^\dagger$ or C$_{11}^\ddagger$ \\
 &  (e) & (kcal/mol/\AA) & (Atm.) &  (kcal/mol) & (\AA$^3$) & (GPa) \\
 &   &  & &  \multicolumn{3}{c}{vdW-DF / ReaxFF} \\ \hline \hline
Cu bcc & - & - & 6.35 & -79.8/-80.1 & 12.0/12.3 & 135.2/116.5 \\
Cu fcc & - & - & 11.42 & -80.7/-81.1 & 11.9/12.4 & 139.5/117.9 \\
Cu sc & - & - & 18.96 & -69.9/-67.6 & 13.9/15.4 & 101.7/104.2 \\ \hline
Mo bcc & - & - & 18.05 & -157.2/-159.1 & 15.8/16.1 & 259.8/189.6  \\
Mo fcc & - & -	 & 21.12 & -148.4/-152.6 & 16.0/17.2 & 237.9/156.5 \\
Mo sc & - & - & 39.39 & -127.3/-124.0 & 17.6/18.2 & 197.2/145.7 \\ \hline
MoCu (CsCl) & 0.06 & 0.00 & 10.57 & X & 26.5/26.3 & 48.2/47.2 \\ \hline
MoTe$_2$ H + 0.125 Cu$_h$ & 0.03 & 0.88 & 2.50 &  27.2/26.2 & 25.8/25.2 & 2.4/2.6$^d$ \\
MoTe$_2$ H + 0.25 Cu$_h$ & 0.03 & 1.20 & 2.64 & 28.4/26.1 & 25.3/24.5 & 2.6/2.6$^d$ \\
MoTe$_2$ H + 0.375 Cu$_h$ & 0.03 & 1.12 & 3.63 &  28.9/25.6 & 24.6/23.8 & 3.4/4.0$^d$ \\
MoTe$_2$ H + 0.5 Cu$_h$ & 0.04 & 1.06 & 4.60 &  29.1/25.1 & 24.0/23.2 & 3.7/4.6$^d$ \\ \hline
MoTe$_2$ H + 0.125 Cu$_t$ & 0.03 & 1.42 & 2.42 &  28.2/28.3 & 25.9/25.7 & 2.4/2.4$^d$ \\
MoTe$_2$ H + 0.25 Cu$_t$ & 0.04 & 1.72 & 1.44 & 30.9/29.5 & 25.6/25.6 & 2.4/3.1$^d$ \\
MoTe$_2$ H + 0.375 Cu$_t$ & 0.04 & 2.18 & 1.50 &  28.8/28.8 & 24.7/24.4 & 3.1/3.7$^d$  \\
MoTe$_2$ H + 0.5 Cu$_t$ & 0.04 & 2.46	 & 2.55 &  24.2/24.7 & 24.9/24.9 & 4.2/5.8$^d$ \\ \hline
MoTe$_2$ H & 0.04 & 1.51 & 5.26 & -31.4/-31.9 & 26.2/26.4 & 45.3/43.6 \\
MoTe$_2$ T & 0.06 & 1.21 & 6.53 & -27.4/-30.0 & 25.8/25.5 & 47.3/58.0 \\
MoTe$_2$ T' & 0.06 & 4.46 & 6.97 & -31.2/-28.6 & 26.5/26.3 & 48.2/47.2 \\ \hline
MoTe$_2$ 3Cu diss & 0.03 & 1.02 & 1.47 & X & X & X\\
MoTe$_2$ diss	& 0.05 &1.18 & 0.70 & - & - & - \\ 
Te$_2$ diss & 0.00 & 2.99 & 0.00 & - & - & - \\ \hline
NEB H$\rightarrow$T' & 0.05 & 2.72 & 2.03 & X  & X & X \\
NEB path1 & 0.02 & 0.49 & 1.35 & - & - & -\\
NEB path2 & 0.02 & 0.56 & 1.29 & - & - & - \\ \hline
\end{tabular}
\end{center}
$^\ast$ RMSE$_i$ = $\sqrt{(i^{\text{vdW-DF}}-i^{\text{ReaxFF}})^2/N_i^2}$ \\
$^\star$ Example for MX$_2$: $E_{f} = E_{MX_2} - (E_M^0+2E_X^0)$ \\
$^\dagger$ From Murnaghan equation \\
$^\ddagger$ From parabola $E = E_0+C_{11}V^2/2$
\label{tab:tab1}
\end{table}%

\subsection{Mechanical properties}

In addition to the energetics analysis, we evaluated the capacity of ReaxFF to describe the mechanical properties of the structures included in the training set.
Equilibrium volumes, bulk moduli and elastic constants are reported on Table \ref{tab:tab1}, computed with both DFT and ReaxFF.
We found between 15 and 52\% error in the description of bulk moduli for bulk Cu and Mo.
However, since the main purpose of the presented force field lies in the description of Cu-intercalated MoTe$_2$, we consider this large error acceptable.
By contrast, ReaxFF describes with great accuracy the moduli corresponding to various MoTe$_2$ phases.
Moreover, ReaxFF describes the gradual stiffening of intercalated MoTe$_2$ with increase number of Cu intercalant, consistent with DFT calculations.

\section{Force field validation: Cu/MoTe$_2$ interface}
\label{sec:app1}

We used the force field to compute the interaction between a single layer of MoTe$_2$ and a (111) Cu surface. 
We follow the strategy developed in Ref.~\cite{komsa2013electronic} to define supercells and minimize strain, chosen to be applied to the metal slab because of the 
corresponding small effect on the electronic structure of the interface~\cite{farmanbar2016first}.
We develop the in-plane supercell lattice vectors $\vec{T}_1=n_1\vec{a}_1+n_2\vec{a}_2$ and $\vec{T}_1^{\prime}=m_1\vec{b}_1+m_2\vec{b}_2$
over the primitive hexagonal cells of MoTe$_2$ \{$\vec{a}_1,\vec{a}_2$\} and the Cu slab \{$\vec{b}_1,\vec{b}_2$\}, respectively. 
From the optimized lattice parameters of monolayer MoTe$_2$ and Cu slab computed with both DFT and ReaxFF, we minimize the lattice mismatch 
$\delta=(|\vec{T}_1|-|\vec{T}_1^\prime|)/|\vec{T}_1|$ and select two interfaces with equivalent strain, as reported on Table \ref{tab:tab2}.

\begin{table}[!ht]
\caption{Parameters corresponding to the supercell construction of a Cu/MoTe$_2$ interface with minimal lattice mismatch ($\delta$).
The two sets ($n_1,n_2$) and ($m_1,m_2$) correspond to the coefficients of the in-plane supercells for MoTe$_2$ and Cu, respectively.
We note that the two supercells have been rotated by the angle $\alpha$.}
\begin{center}
\begin{tabular}{c|cccc}
Method & $n_1,n_2$ & $m_1,m_2$ & $\alpha$ ($^\circ$) & $\delta$ (\%) \\ \hline \hline
ReaxFF & 5,4 & 7,5 & 1.8 & 1.5 \\
vdw-DF & 2,1 & 3,1 & 5.2 & 1.3 \\ \hline
\end{tabular}
\end{center}
\label{tab:tab2}
\end{table}%

\begin{figure}[!ht]
\centering
\includegraphics[width=0.45\textwidth]{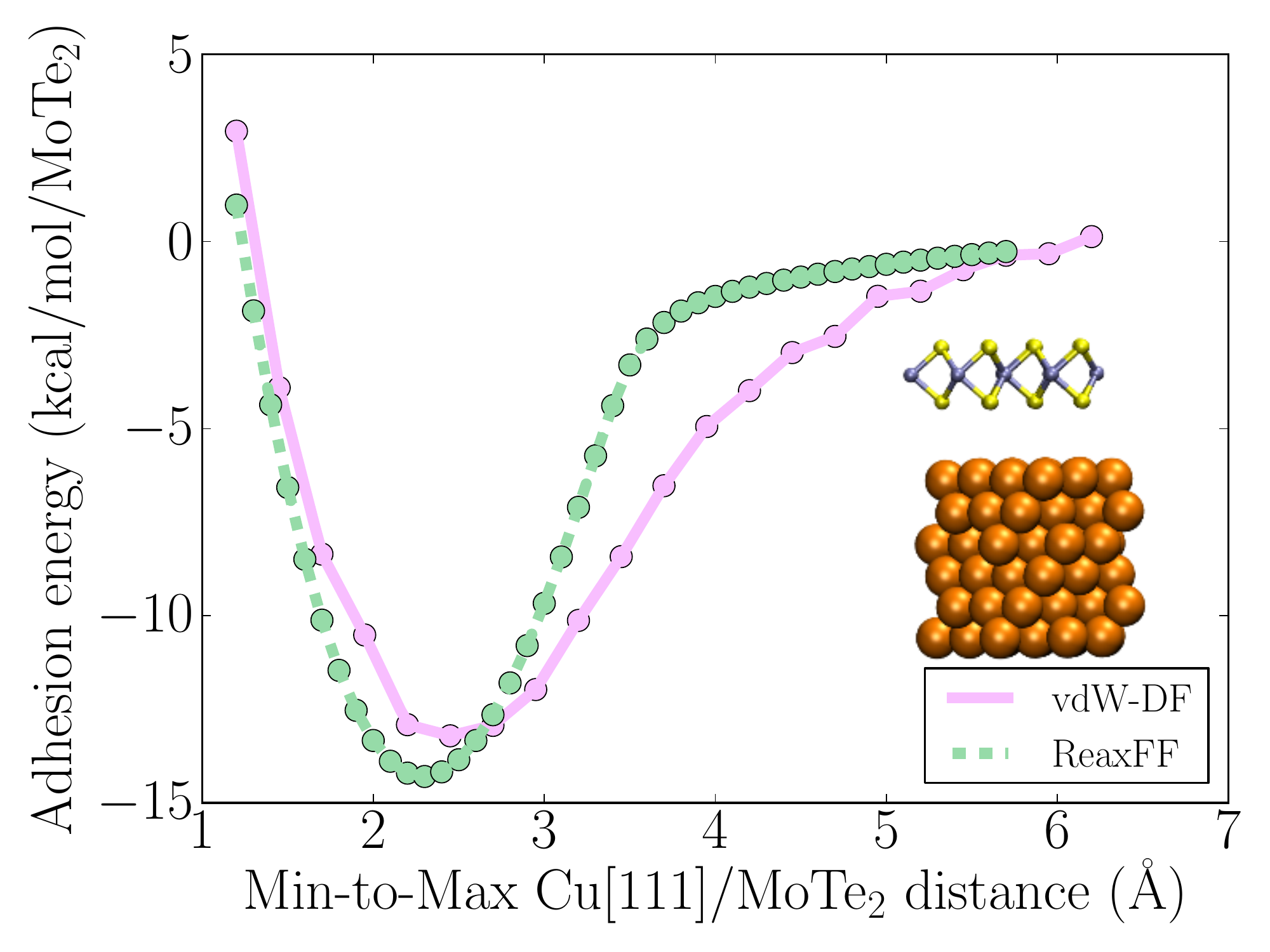}
\caption{Adhesion energy as a function of the minimum separation between a slab of Cu(111) and MoTe$_2$ computed with vdW-DF (solid) and ReaxFF (dashed).}
\label{fig:fig8}
\end{figure}

Figure \ref{fig:fig8} shows the adhesion energy (the energy of the interface minus the energy of the Cu slab minus the energy of the MoTe$_2$ layer) 
computed with DFT and ReaxFF.
We find very good agreement between the force field and DFT calculations even though such data was not included in the force field parameterization.
This indicates the type of regimes the force field is expected to be extrapolated to. However, as with any force field, care should be exercised when using
it away for the conditions for which it was trained.

\section{Diffusion of Cu-intercalated MoTe$_2$}
\label{sec:app2}

The relatively small computational intensity of ReaxFF enables the simulations of large scale systems; providing a unique tool for exploring 
ionic transport, fracture and the effect of defects on mechanical properties and growth. In this section we focus on the first of these examples.

We know from the first principle calculations presented above that the activation energy for thermal hop of Cu intercalated MoTe$_2$ varies between 9-13 kcal/mol 
(0.39-0.56 eV) range, depending on the intercalation site. These values are intermediate between the extreme cases of a good ion conductor like highly metal doped 
chalcogenide glasses with E$_A\sim$ 0.2 eV~\cite{soni2010probing} and amorphous silica with in average E$_A\sim$ 0.8 eV~\cite{guzman2015stability}. 
In order to confirm this last statement, we performed simulations of Cu-intercalated bulk MoTe$_2$ at various temperature ranging from 
340 to 380K and evaluated the diffusion coefficient D $\approx$ [10$^{-7}$-10$^{-8}$] cm$^2$/s (see Supplementary Materials). 
Therefore, within the timescales of MD and at room temperature we find negligible thermal diffusion; thus we study the drift of the ions under an external bias, 
relevant to estimate the electromobility of Cu-intercalated ions in MoTe$_2$-based device operation.

\begin{figure}[!ht]
  \centering
  \includegraphics[width=0.45\textwidth]{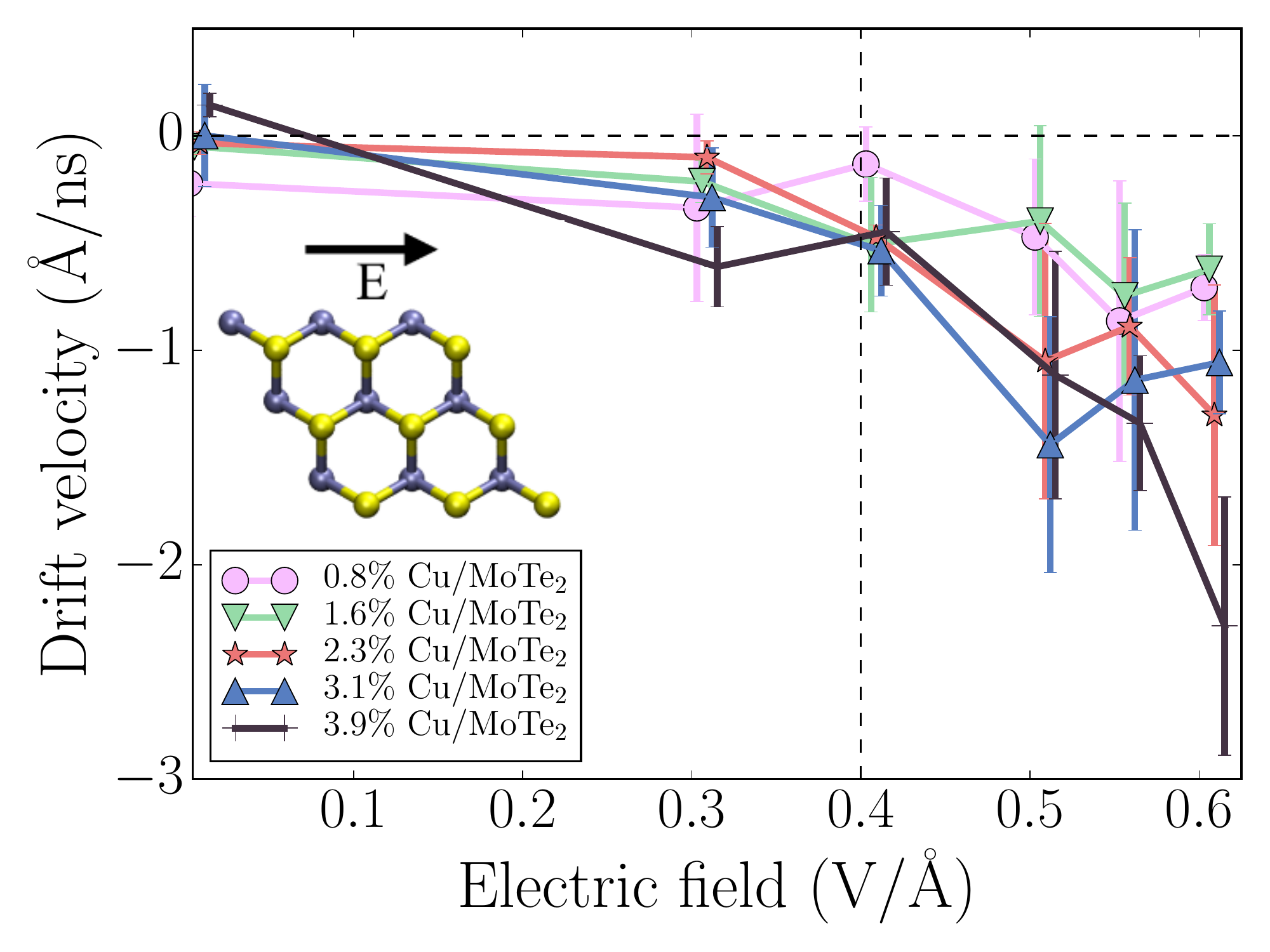}
    \includegraphics[width=0.45\textwidth]{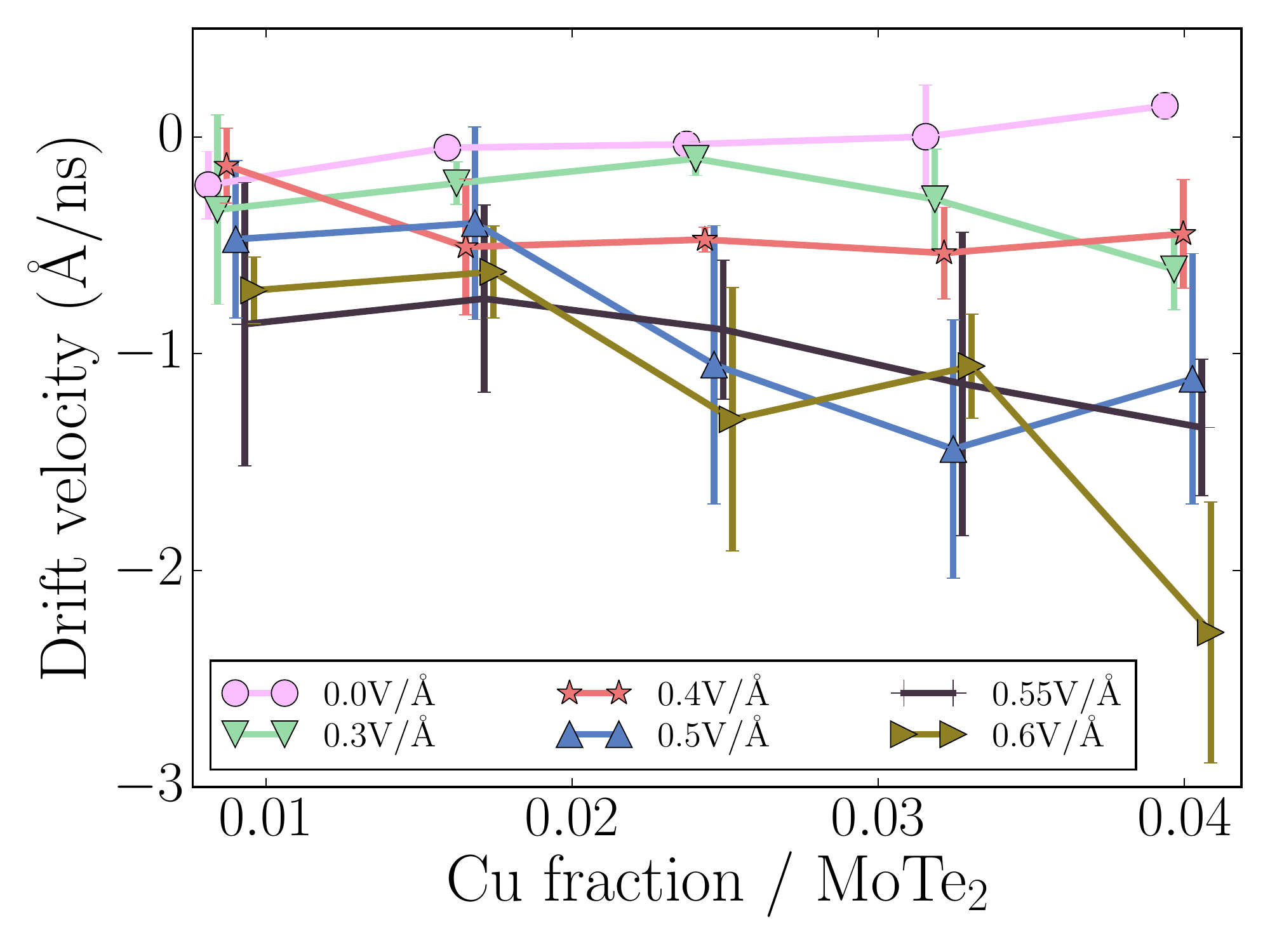}
  \caption{Drift velocity of Cu ions as a function of the electric field for various Cu fraction (left) and as a function of the Cu fraction for various electric field (right).
The error bars have been evaluated from the standard deviation of drift velocity over 4 independent samples and the data points between curves
  have been slightly shifted in order to visualize the error bars.}
  \label{fig:fig9}
\end{figure}

We now explore the effect of Cu concentration on ion mobility in Cu-intercalated bulk MoTe$_2$ under the action of an external electric field.
We performed MD simulations of Cu-intercalated bulk H-MoTe$_2$ in the canonical ensemble at 300 K for up to 2 ns and applied various external electric fields
(i.e. a constant force F$_i$ = Eq$_i$ to each atom $i$) ranging from 0.0 to 0.6 V/\AA. 
The direction of the electric field is chosen to be aligned with the in-plane lattice parameter of the hexagonal cell of MoTe$_2$, as shown in the inset 
of Figure \ref{fig:fig9}.
We varied the concentration of Cu ions from 1 to 4\% with respect to MoTe$_2$ and study their mobility.
In order to estimate uncertainties in the mobility of the Cu ions, we performed simulations over an ensemble of 4 independent samples differing
by the initial position and velocity of copper atoms.

Since thermal diffusion leads to no net  motion of the atoms, the average position of the Cu ions as a function of time $<X(t)>$, can be described as:

\begin{equation}
<X(t)> = X_0 + v_dt
\end{equation}

The mobility $u$ can then be obtained from the applied electric field $E$ and drift velocity $u=v_d/E$~\cite{ren2014structural}. 
Figure \ref{fig:fig9} shows the computed drift velocity of Cu-intercalated MoTe$_2$ as a function of the electric field (left) and the Cu fraction (right). 
We found that a minimum electric field of approximately 0.4 V/\AA~is required to displace the Cu ions in the vdW gap of MoTe$_2$ within MD timescales. 
More importantly, we find that the ionic mobility increases with concentration. Finally, since ReaxFF predicts slightly higher diffusion barriers, with respect to 
DFT calculations, we expect the electric field threshold for drift diffusion to be lower than the 0.4 V/\AA~predicted here.

\section{Conclusions}
\label{sec:conc}

We presented a parameterization of ReaxFF to describe the interactions between MoTe$_2$ and copper and showed that such force field can be used to study
large scale atomistic simulations of interest for emerging electronics.
The optimization of the force field against first principle calculations shows that ReaxFF reproduces with accuracy the energetics, charges and mechanical properties 
of various systems composed of MoTe$_2$ and Cu and slightly overestimates the energy barrier for Cu diffusion in the vdW gap of the TMD.
Additionally, we demonstrated the good transferability of the force field to describe Cu/MoTe$_2$ interfaces.
As for any force field, care must be taken when used outside of the conditions used during parameterization.
For example, the force field has not been optimized to describe defect formation energies and extensive validations must be performed in order to study growth, as we
envisage in the future.

\FloatBarrier
\subsection*{Acknowledgement}
We acknowledge Benjamin Helfrecht for DFT calculations of the binding energy between Cu and MoTe$_2$. This work was partially supported by the FAME and LEAST Centers, two of six centers of STARnet, a Semiconductor Research Corporation program sponsored by MARCO and DARPA. We thank nanoHUB.org and Purdue for the computational resources.
 
\clearpage

\end{document}